\documentclass[twocolumn]{aastex7}
\usepackage{mhchem}
\usepackage{leftindex}
\usepackage[colorinlistoftodos]{todonotes}
\usepackage{amssymb}
\usepackage{comment}
\usepackage{multirow,acronym}
\usepackage{natbib}
\usepackage{listings} 
\usepackage{makecell}
\usepackage{booktabs}
\usepackage{amsmath}
\usepackage{subfigure}
\usepackage{color}
\usepackage{xcolor}
\usepackage{hyperref}
\usepackage[T1]{fontenc}
\usepackage{graphicx}
\usepackage{bm}
\hypersetup{colorlinks=true, citecolor=blue, 
  linkcolor=cyan, urlcolor=magenta} 
\usepackage{aas_macros}
\usepackage{amsfonts}
\usepackage{float}
\usepackage{amsmath,amssymb}
\usepackage{natbib}    
\usepackage{hyperref} 
\usepackage{graphicx} 
\usepackage{color}
\usepackage{verbatim}
\usepackage{enumitem}
\usepackage{natbib}

\usepackage[linesnumbered,ruled]{algorithm2e}

\newcommand{\proptosim}{\mathrel{\vcenter{
 \offinterlineskip\halign{\hfil$##$\cr
 \propto\cr\noalign{\kern2pt}\sim\cr\noalign{\kern-2pt}}}}}
\renewcommand{\b}[1]{\boldsymbol{#1}}
\newcommand{\unit}[1]{{\rm #1}}

\renewcommand{\min}{\mathrm{min}}
\renewcommand{\max}{\mathrm{max}}
\newcommand{\au}{\mathrm{AU}}
\newcommand{\cm}{\unit{cm}}
\newcommand{\m}{\unit{m}}
\newcommand{\g}{\unit{g}}
\newcommand{\G}{\unit{G}}
\newcommand{\K}{\unit{K}}
\newcommand{\km}{\unit{km}}

\renewcommand{\micron}{\unit{\mu m}}

\newcommand{\eV}{\unit{eV}}

\newcommand{\s}{\mathrm{s}}
\newcommand{\yr}{\mathrm{yr}}

\newcommand{\ang}{\text{\AA}}

\newcommand{\B}{\mathbf{B}}     
\newcommand{\E}{\mathbf{E}}     
\newcommand{\J}{\mathbf{J}}     
\renewcommand{\v}{\mathbf{v}}   
\newcommand{\kb}{k_\mathrm{B}}

\renewcommand{\d}{\mathrm{d}}

\newcommand{\e}{\mathrm{e}}

\newcommand{\p}{\mathrm{p}}     

\newcommand{\eff}{\mathrm{eff}}

\newcommand{\crit}{\mathrm{crit}}

\renewcommand{\O}{\mathrm{O}}     
\newcommand{\A}{\mathrm{A}}     
\renewcommand{\H}{\mathrm{H}}     
\renewcommand{\P}{\mathrm{P}}     
\newcommand*\chem[1]{\ensuremath{\mathrm{#1}}}

\newcommand{\br}{\mathrm{br}}
\newcommand{\jade}{\texttt{JADE}}
\newcommand{\figdir}{.}

\newcommand{\DOA} {Department of Astronomy, School of
  Physics, Peking University, Beijing 100871, China}
\newcommand{\KIAA}{Kavli Institute for Astronomy and
  Astrophysics, Peking University, Beijing 100871, China}
 
\begin{document}

\title{Consistent Modeling of Non-equilibrium Dust
  Sublimation and the Interactions with \\
  Dust Evolution in the Inner Regions of Protoplanetary
  Disks}

\author[0000-0003-2895-4968]{Sheng Xu}
\affil{\DOA}
\affil{\KIAA}
\email{sx?????@stu.pku.edu.cn}

\author[0000-0002-6540-7042]{Lile Wang}
\affil{\DOA}
\affil{\KIAA}
\email{lilew@pku.edu.cn}

\author[0000-0001-6947-5846]{Luis C. Ho}
\affiliation{\KIAA}
\affiliation{\DOA}
\email{lho.pku@gmail.com}

\author[0000-0001-8531-9536]{Renyue Cen}
\affiliation{Institute for Advanced Study in Physics,
  Zhejiang University, Hangzhou 310027, China}
\affiliation{Institute of Astronomy, School of Physics,
  Zhejiang University, Hangzhou 310027, China}
\email{renyuecen@zju.edu.cn}

\author[0000-0001-7268-9917]{Shenzhen Xu}
\affil{School of Material Sciences,
  Peking University, Beijing 100871, China}
\email{szxu@pku.edu.cn}

\correspondingauthor{Lile Wang}
\email{lilew@pku.edu.cn}

\begin{abstract}
  The inner regions of protoplanetary disks are host to the
  sublimation of dust grains, a process traditionally
  modeled using equilibrium thermodynamics. We demonstrate
  through {\it ab-initio} density functional theory (DFT)
  and kinetic Monte Carlo (KMC) simulations that silicate
  dust sublimation is inherently a non-equilibrium kinetic
  process. The binding energies and vibrational frequencies
  governing desorption, calculated for \chem{MgSiO_3} and
  other compositions, reveal that sublimation timescales far
  exceed local dynamical times, allowing grains to persist
  in a superheated state. This kinetic inhibition results in
  a broad, dynamic sublimation front whose location and
  morphology are strongly regulated by radial advection and
  dust coagulation. Our coupled simulations, integrating
  sublimation with advection and grain evolution, show that
  the front varies radially by a factor of four with
  accretion rate and exhibits a vertically stratified,
  bowl-shaped structure. These findings imply that the inner
  disk dust distribution, thermal structure, and subsequent
  planet formation are profoundly influenced by the
  kinematics and kinetics of dust grains, necessitating a
  departure from equilibrium prescriptions in disk models
  and interpretations of inner rim observations.
\end{abstract}


\section{Introduction}
\label{sec:intro}

Dust grains are the foundation upon which planetary systems
are assembled.  In the innermost regions of protoplanetary
disks (PPDs), where temperatures exceed $\gtrsim 1000~\K$,
the first solids to disappear are believed to be the
silicates that dominate the solid-mass budget at $0.1-2~\au$
\citep[e.g.][]{2005A&A...438..899I} Their removal fixes the
inner edge of the dust disk, truncates the radial-drift flow
that feeds terrestrial-planet formation, and determines the
chemical boundary conditions for the accretion of hot,
rock-forming vapour onto young gas giants \citep[see also]
[and references therein]{2010ARA&A..48..205D}. Recent
studies on the accretion mechanism have also emphasized the
key role of dust grains in maintaining intermediate coupling
between magnetic fields and materials, which is a necessity
for the wind-driven laminar accretion process whose
efficiency overwhelms turbulent viscous accretion mechanism
\citep{2009ApJ...701..737B, 2016ApJ...819...68X,
  2017ApJ...845...75B, 2019ApJ...874...90W}. Vaporized dust
grains could lead to a significantly lowered accretion rate
and reduced opacity, shaping an inner rim near the
sublimation front via various mechanisms
\citep{2012MNRAS.420.1541V, 2016ApJ...827..144F}.

While the macroscopic importance of dust grain in the PPD
inner regions emerges from both theories and observations,
the detailed sublimation mechanism has been treated as an
equilibrium process in most models. A grain is assumed to
sit at the local equilibrium temperature, with surface
vapour pressure taken to be the saturation
value. Instantaneous sublimation equilibrium criterion is
adopted: dust survives where $T<T_{\rm sub}(P_{\rm gas})$
and vanishes where $T\ge T_{\rm sub}(P_{\rm gas})$
\citep[see also][]{1994ApJ...421..615P,
  2005A&A...438..899I}.  This simplification is embedded
into dust-evolution models in theories and simulations, and
retrieval of inner-disk radii from observations including
infrared interferometry \citep[see e.g.][]{
  1994ApJ...427..987B, 2013ChRv..113.9016H,
  2016SSRv..205...41B, 2022A&A...658A..36K}.

Deviations from the sublimation-condensation equilibria,
nevertheless, have started to propagate into macroscopic
observables.  At radii where equilibrium models predict a
sharp dust wall, non-equilibrium sublimation can smear the
transition over a radial width
$\Delta r\simeq 0.05-0.08~\au$, which have been attributed
to the turbulent diffusion or accretion
\citep[e.g.][]{2021A&A...651A..27S}.  High-resolution
infrared spectra reveal that the $10~\micron$ silicate
feature weakens inside $\sim 0.3~\au$ in typical T Tauri
disks \citep{2006ApJ...639..275K}, while interferometric
continuum visibilities locate the inner rim at radii a
factor of two smaller than predicted by equilibrium
condensation temperatures. In addition, the corresponding
infrared continuum slope is shallower and time-variable on
the scale of weeks, in qualitative agreement with recent
VLTI/GRAVITY monitoring of the Herbig Ae star HD 163296
(\citealt{2021A&A...654A..97G}; see also
\citealt{2023A&A...669A..59G}).  Attempts to reconcile these
discrepancies have invoked porosity compositional gradients,
back-warming by the stellar magnetosphere, or optically
thick wall geometries \citep{2001ApJ...560..957D,
  2003ApJ...591.1220L, 2008ApJ...689..513T,
  2012MNRAS.420.1541V, 2015A&A...582L..10K}.

More sophisticated models coupled with macroscopic
astrophysics could reveal more physical mechanisms related
to grain sublimation. Vapour produced at the sublimating
surface is injected into the gas phase at supersaturations
of $10^{2}-10^{3}$, providing a natural trigger for
homogeneous nucleation of silica-rich nanoparticles
\citep{2004A&A...413..571G}.  These freshly condensed grains
are optically thin and dynamically coupled to the gas,
offering an explanation for the persistent NIR excess and
the extreme depletion of refractory elements in the
accretion columns of T Tauri stars
\citep{2015A&A...582L..10K}.  \citet{2024PASJ...76..616I}
discussed the issue of the inner boundary of protoplanetary
disks, emphasizing that a more detailed consideration of the
dust sublimation process could offer new insights into the
extent of this boundary. The sublimation rate is a function
of temperature, and over the typical protoplanetary disk
lifetime of around $\sim 10^6~\yr$, the slow sublimation of
dust at lower temperatures can release alkali and alkaline
earth metals. These released species influence the
ionization degree of the disk, thereby affecting the
truncation boundary of magneto-rotational instability (MRI).

To quantify the kinetics of grain sublimation, experimental
studies have been carried out, whose effective temperature
range is nevertheless significantly lower than the
sublimation conditions in the inner disks \citep[e.g.][]
{2022MNRAS.509.2825S}. In order to cover the thermodynamic
conditions relevant to inner-disk astrophysics, we conduct
the {\it ab-initio}, atomistically resolved study of sublimation
processes that follows the actual kinetics of bond breaking,
diffusion, and desorption using density-functional theory
(DFT) and kinetic Monte-Carlo (KMC) simulations at the
surfaces of dust grains. This paradigm have already been
extensively adopted in the studies of material properties
regarding their sublimation processes \citep[e.g][]
{2023ACSNa..17.8098H}. Such calculations could
quantitatively model sublimation as non-equilibrium
processes, whose characteristic time-scale can be comparable
or even exceed the dynamical time of the disk. 
Admittedly, this approach focuses crystalline materials,
while the extension to amorphous materials is delayed to
following works. We nevertheless note that the simulations
for crystals are still valid for polycrystalline grains, and
the physical quantitites obtained are expected to hold
semi-quantitatively in preliminary discussions on amorphous
materials. Combined with proper models of dust coagulation
and fragmentation, resolving sublimation physics is not only
a refinement of modeling inner regions of PPDs, but could
also help to develop further insights into the in-situ
formation of rocky planets and planetesimals close to the
central protostar.

This paper is structured as follows. \S\ref{sec:micro}
presents the microscopic core of the study, including the
DFT calculations of atom-specific binding
energies and kinetic Monte-Carlo simulations that expose
sublimation as an intrinsically non-equilibrium,
surface-activated process whose time-scales can rival
orbital periods.  Building on these microphysics,
\S\ref{sec:apps-dust-survival} follows the fate of dust
grains in a one-dimensional advection model and shows that
survival is governed by the competition between kinetically
limited mass loss and radial replenishment.
\S\ref{sec:apps-sub-2d} then embeds the kinetic rates in a
2.5-D framework where coagulation, fragmentation and
vertically stratified accretion are co-evolved; this reveals
a broad, bowl-shaped sublimation front whose location spans
0.05–0.15 au and whose structure is sculpted by accretion
rate and grain growth.  Finally, \S\ref{sec:discussion}
synthesises these findings, discusses caveats arising from
prescribed velocity fields and dust-to-gas coupling, and
charts future work.

\section{Simulations for Sublimation}
\label{sec:micro}

To accurately simulate the sublimation of silicate dust
grains, the model should encompass a sufficiently large
surface area to properly represent the anisotropic escape of
atoms from different crystal surfaces and the evolution of
surface morphology. Conducting full-scale {\it ab-initio}
molecular dynamics (AIMD) simulations at the quantum
mechanical level (e.g., using DFT) is computationally
prohibitive due to the immense number of atoms required to
represent a realistic grain surface and the long timescales
of the sublimation process. Given the fact that
astrophysical dust grains are mostly in stiff equilibrium
with the diffuse radiation fields originating from the
centro protostar \citep[see][]{1997ApJ...490..368C}, we
adopt a two-step multi-scale simulation strategy that
bridges the accuracy of quantum mechanics with the
scalability KMC methods, elaborated in what follows.

\subsection{{\it Ab-initio} Calculations for the Interactions}
\label{sec:micro-dust-dft}

\begin{figure}
  \centering
  \includegraphics[width=0.95\linewidth]{\figdir/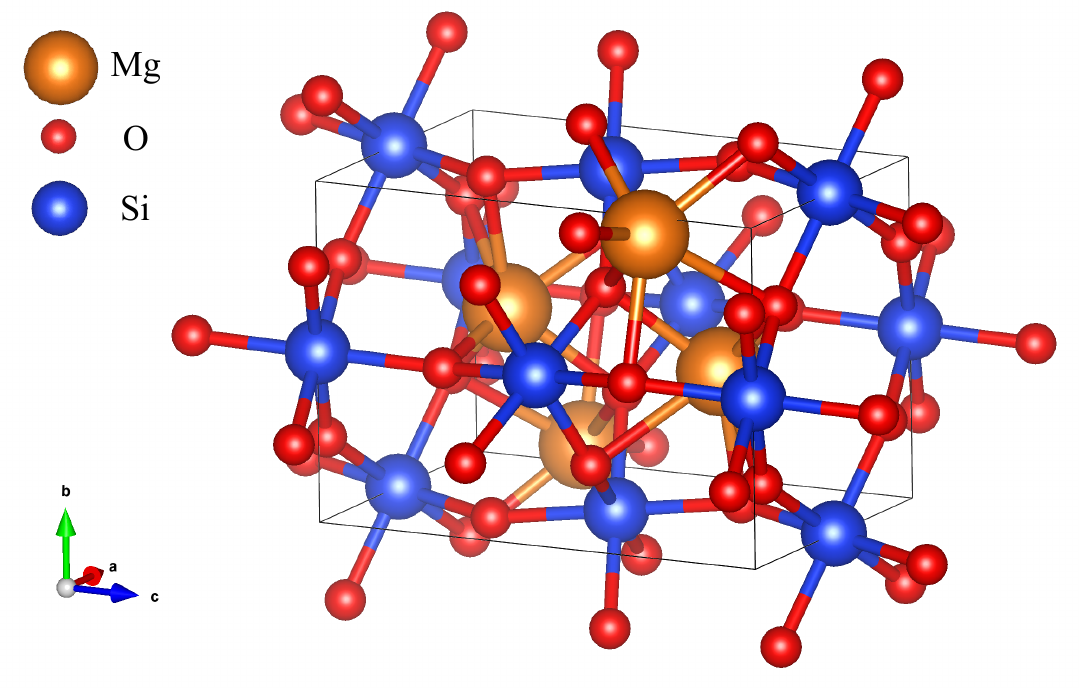}
  \caption{Structure of the model silicate \chem{MgSiO_3},
    showing a cell in the \texttt{pnma} space group 
    \citep[adopted from the Materials Project;][]{MatProj}. }
  \label{fig:lattice-pnma}
\end{figure}

The accuracy of sublimation simulations depend upon the
precise determination of the interatomic interaction
energies within the solid phase. To achieve this, we employ
DFT, the de facto standard for first-principles electronic
structure calculations in molecules and condensed matter
systems. Our detailed DFT computations were performed using
the Vienna Ab initio Simulation Package (VASP;
\citealt{kresse1996efficiency}), which is widely adopted in
modeling solid-state and surface properties.  As the
complication and diversity of silicates could be prohibitive
for detailed and thorough microscopic modeling, several
observational constraints have pointed out that enstatites
(\chem{MgSiO_3}) can be adopted as a reasonable and
representative proxy for magnesium-rich astronomical
silicates \citep[e.g.][] {2002A&A...382..222M,
  2002A&A...382..241M}. For our fiducial silicate dust grain
model, we selected a crystalline structure of magnesium
metasilicate ($\chem{MgSiO_3}$) with the orthorhombic
perovskite structure in the \texttt{pnma} space group (also
know as bridgmanite) from the Materials Project
\citep{MatProj}, illustrated in
Figure~\ref{fig:lattice-pnma}. The binding energy was
computed as the difference in the total energy of the system
between two states: (1) the pristine, fully relaxed lattice,
and (2) the same system where the target atom is displaced
to a distance greater than $9~\ang$ from the bulk surface,
effectively eliminating its chemical interactions. In the
second configuration, the remaining atoms in the bulk phase
are allowed to relax to their new energy-minimized positions
to ensure a consistent and physically accurate energy
comparison. We have also confirmed with DFT-based energy
curves for the interaction energy as a function of atom-bulk
distance (not illustrated in this paper), that there are no
extra surface potential barriers raising the activation
energy of sublimation. Such absence of extra barrier
guarantees that the binding energy values can be directly
adopted by the KMC procedures for sublimation processes.

It is critical to note that the binding energy of an atom is
not a fixed value but is highly sensitive to its local
coordination environment. This environment is defined by
both the number (coordination number) and chemical
identities of its nearest neighbors. Consequently, we
systematically calculated the binding energies for Mg, Si,
and O atoms across a variety of coordination states. In the
fully coordinated bulk crystal, Mg and Si atoms are
typically octahedrally and tetrahedrally coordinated by six
and four oxygen atoms, respectively, while oxygen atoms are
bonded to two Mg and two Si atoms. The binding energies for
various fully-coordinated and under-coordinated scenarios
(e.g., surface or defect sites) are catalogued in
Table~\ref{table:ene-mgsio3}. For coordination environments
not explicitly listed, values were estimated via linear
interpolation based on the number and type of missing
neighbors, providing a complete energy landscape for our
kinetic model.

To ensure a consistent foundation for our KMC simulations,
the characteristic vibrational eigenfrequencies of atoms
(essential for calculating attempt frequencies for
desorption) were derived from a numerical evaluation of the
Hessian matrix within VASP. Furthermore, we justify the
neglect of van der Waals (vdW) interactions in our
calculations by their secondary energy contribution (on the
order of $10^{-1}~\eV$) dwarfened by the values and inherent
error of the much stronger ionic-covalent bonding energies
(on the order of several eV) governing silicate
sublimation. Their exclusion thus significantly enhances
computational efficiency without compromising the
quantitative accuracy of our results.

\begin{deluxetable}{ccc}
  \tablecolumns{3} 
  \tabletypesize{\scriptsize}
  \tablecaption{Binding energy of atoms in model silicates
    (\chem{MgSiO_3}). \label{table:ene-mgsio3} }
  \tablehead{
    \colhead{Concerned atom} &
    \colhead{Nearest neighbors} &
    \colhead{$\Delta E/\eV$}
  }
  \startdata
  Mg & 6O & 11.24 \\
     & 5O & 3.88 \\ 
     & 1O & 3.01 \\
  \hline
  Si & 6O & 19.85 \\ 
     & 5O & 11.93 \\
     & 1O & 5.37 \\
  \hline
  O & 2Mg + 2Si & 9.75 \\
    & 2Mg + 1Si & 1.81 \\
    & 1Mg + 2Si & 2.44 \\
    & 1Mg       & 3.72 \\
    & 1Si & 3.17 \\
  \enddata
\end{deluxetable}

\subsection{KMC Simulations for Sublimation}
\label{sec:micro-kmc}

Using the comprehensive binding energy and vibrational
frequency data obtained via the DFT methods described in
\S\ref{sec:micro-dust-dft}, we perform KMC simulations to model
the non-equilibrium sublimation process of $\chem{MgSiO_3}$
dust grains. KMC simulates possible surface reactions,
diffusion hops, and desorption events
\citep[e.g.][]{2019FrCh....7..202A}, allowing us to
transcend the limitations of equilibrium thermodynamics by
explicitly tracking the stochastic desorption and surface
diffusion events that occur over macroscopic timescales. The
simulation is initialized by constructing a
three-dimensional lattice representative of a specific
crystal surface with a sufficiently large bulk phase beneath
it to accurately model the energetic environment of
sub-surface atoms. Then, the KMC algorithm proceeds
iteratively as follows:
\begin{enumerate}
\item Catalog all possible events and their rates: For every
  atom on the surface and in the sub-surface layers, all
  possible events $i$ are identified. These primarily
  include (1) desorption into the vacuum, and 
  (2) surface diffusion to a neighboring vacant site.
\item Evaluate the rate $r_i$ for each event is calculated
  using $r_i = \nu_i \exp( -E_{\mathrm{a},i}/\kb T)$, where
  $\nu_{i}$ is the vibrational attempt frequency (obtained
  from DFT Hessian matrix calculations), $E_{a,i}$ is the
  activation energy for the event (the binding energy or
  diffusion barrier), and $\kb$ is the Boltzmann
  constant. We assume that the activation energy of an
  atomic dissociation equals to the corresponding binding
  energy, indicating no extra barrier along the dissociation
  process.
\item Calculate the total rate for all $N$ possible events
  by summing $R_{\rm tot} = \sum_{i=1}^{N} r_i$.  
\item A cumulative list of rates is constructed, where the
  $n$-th event corresponds a cumulative sum
  $S_n = \sum_{i\leq n} r_i / R_{\rm tot}$.  Select an event
  to execute by generating a uniformly distributed random
  number $X\in [0, 1]$: the event $m$ is chosen such that it
  satisfies the condition $S_{m} \leq X < S_{m+1}$.
\item Advance the simulation clock 
  whose timestep is set stochastically according using
  $\Delta t = -\ln Y/R_{\rm tot}$ ($Y\in [0, 1]$ is another
  uniformly distributed random number).
\end{enumerate}
These steps, which are ordinarily adopted in most other KMC
simulations, are repeated for sufficient number of
iterations, allowing the simulation to track the evolution
of the grain surface morphology and the rate of mass loss
(sublimation) over physically significant timescales.  Such
algorithm efficiently captures the kinetic evolution of the
dust grain surface, enabling the direct simulation of
sublimation rates under conditions where the system is far
from thermodynamic equilibrium.

\begin{deluxetable}{cccc}
  \tablecolumns{4} 
  \tabletypesize{\scriptsize}
  \tablecaption{Fitting parameters for the sublimation rates
    in the Boltzmann function form
    (eq.~\ref{eq:sub-boltzman}). \label{table:fit-boltzman} }
  \tablehead{
    \colhead{Component} &
    \colhead{Surface index} &    
    \colhead{$\ln S_{i0}$} &
    \colhead{$\Delta E_{i}/\kb$} \\
    \colhead{} &    
    \colhead{} &
    \colhead{$({\rm g\ cm}^{-2}{\rm s}^{-1})$} &
    \colhead{$(10^4~\K)$}
  }
  \startdata
  \chem{MgSiO_3} & (100) & 18.10 & 4.328 \\
  & (010) & 17.65 & 4.217 \\
  & (001) & 21.22 & 10.036 \\  
  \hline
  \chem{MgFeSiO_4} & (100) & 20.06 & 8.478 \\
  & (010) & 16.29 & 7.442 \\
  & (001) & 22.23 & 9.059 \\    
  \hline
  Graphene & - & 12.66 & 13.009 \\    
  \enddata
\end{deluxetable}

\begin{figure}
  \centering
  \includegraphics[width=1.0\linewidth]
  {\figdir/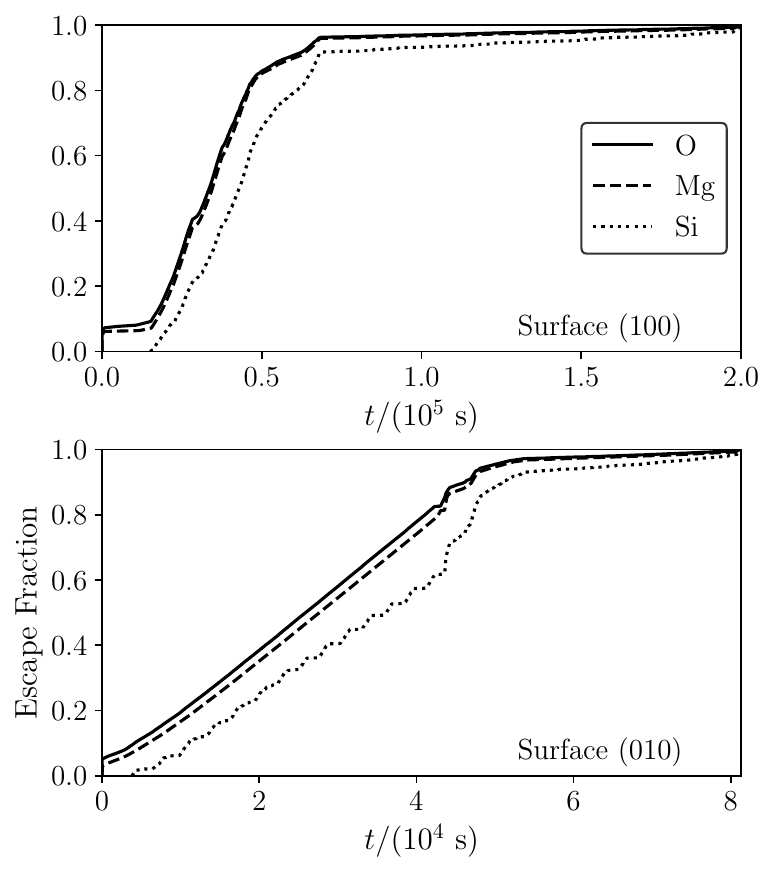}
  \caption{Example of the statistics in KMC simulations on
    the sublimation of \chem{MgSiO_3}, showing the fraction
    of escaped atoms to the total number in the bulk as the
    sublimation history over different crystal surfaces
    (indicated in each panel) at $T = 10^3~\K$. Escaping
    atom types are distinguished by line styles. Note that
    the timescales are very different across the panels, and
    the $(001)$ surface is not illustrated as the escape
    rate is excessively low (escape timescale
    $\gtrsim 10^{28}~\s$) due to the Boltzmann factor given
    by high confinement energy relative to $10^3~\K$ (see
    also Table~\label{table:fit-boltzman}). }
  \label{fig:kmc-mgsio3}
\end{figure}

\begin{figure}
  \centering
\includegraphics[width=1.0\linewidth]
{\figdir/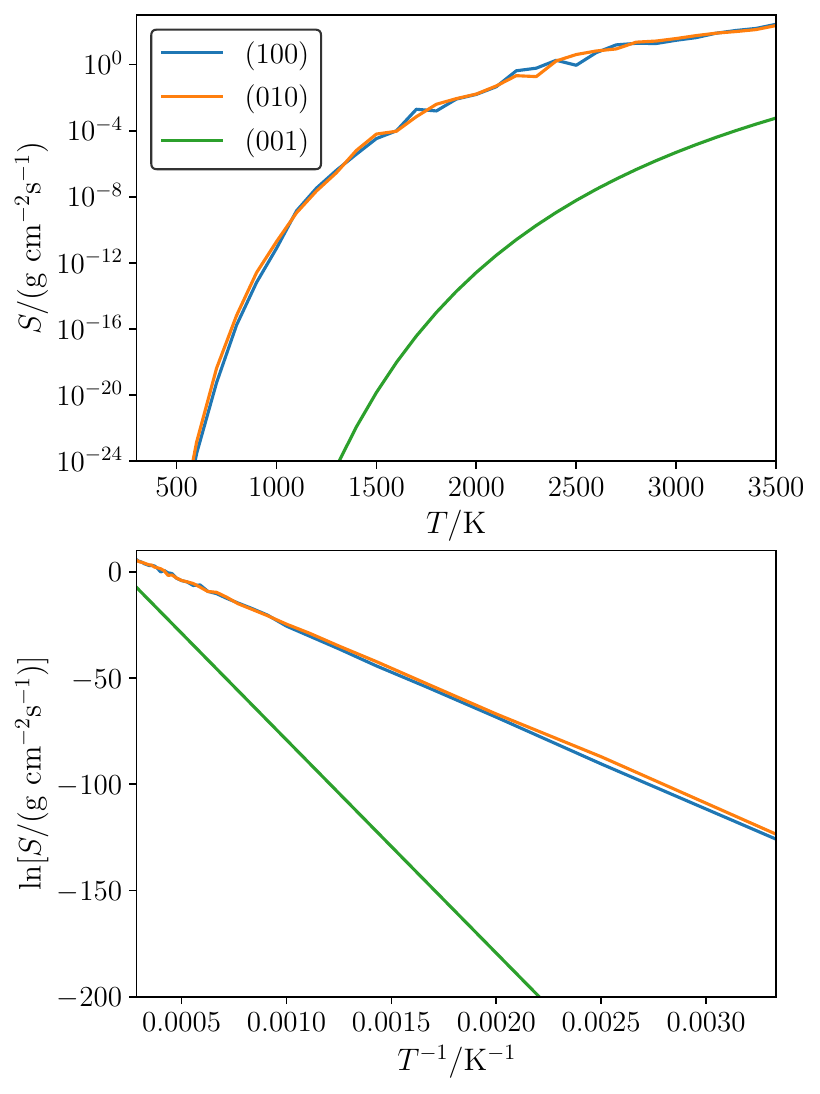}
\caption{Sublimation rates per unit surface area of
  \chem{MgSiO_3} over three representative surfaces,
  distinguished by line colors. While the upper panel
  exhibits in the normal form, the lower presents the
  $\ln S - T^{-1}$ relation to illustrate the fitting on the
  Boltzmann function (see also eq.~\ref{eq:sub-boltzman} and
  Table~\ref{table:fit-boltzman}). }
\label{fig:rate-mgsio3}
\end{figure}

Using a bulk phase with $10^4$ atoms, the sublimation
history of an example simulation on the \chem{MgSiO_3}
silicate model is illustrated in
Figure~\ref{fig:kmc-mgsio3}. Cumulating the history of atom
escape events, one can obtain the sublimation rate over
different crystal surfaces presented in
Figure~\ref{fig:rate-mgsio3} as functions of
temperatures. It is observed that the sublimation rates per
unit surface area of surfaces (100) and (010) are similar,
being significantly greater than (001) by several orders of
magnitude. Such phenomenon is attributed to the fact that
the (001) surface has alternating layers of magnesium and
silicon atoms, which could maximize the binding as silicon
atoms exhibit high binding energy even at low coordination
numbers. Figure~\ref{fig:rate-mgsio3} also includes a panel
plotting the $\ln S-T^{-1}$ relations, easing the fitting
towards the Boltzmann function to approximate the
sublimation rate through the crystal surface indexed as $i$,
\begin{equation}
  \label{eq:sub-boltzman}
  S_i = S_{i0} \exp\left(-\dfrac{-\Delta
      E_i}{\kb T}\right)\ .
\end{equation}
The parameters for the three representative crystal surfaces
of \chem{MgSiO_3} are summarized in
Table~\ref{table:fit-boltzman}. In practice, when the
actual sublimation rates are desired in the co-evolution
calculation of dust evolution and sublimation, the average
over possible crystal surfaces are taken to model the
realistic situations.

\subsection{Exploration on Dust Compositions}
\label{sec:micro-others}

In addition to the fiducial magnesium silicate
($\chem{MgSiO_3}$) model, our simulation framework is
extended to investigate a range of other chemically distinct
dust grains. This exploration can be useful in constructing
a comprehensive understanding of the inner disk environment,
where a diverse population of solids coexists and evolves
under intense thermal radiation.

\begin{figure}
  \centering
  \includegraphics[width=0.85\linewidth]
  {\figdir/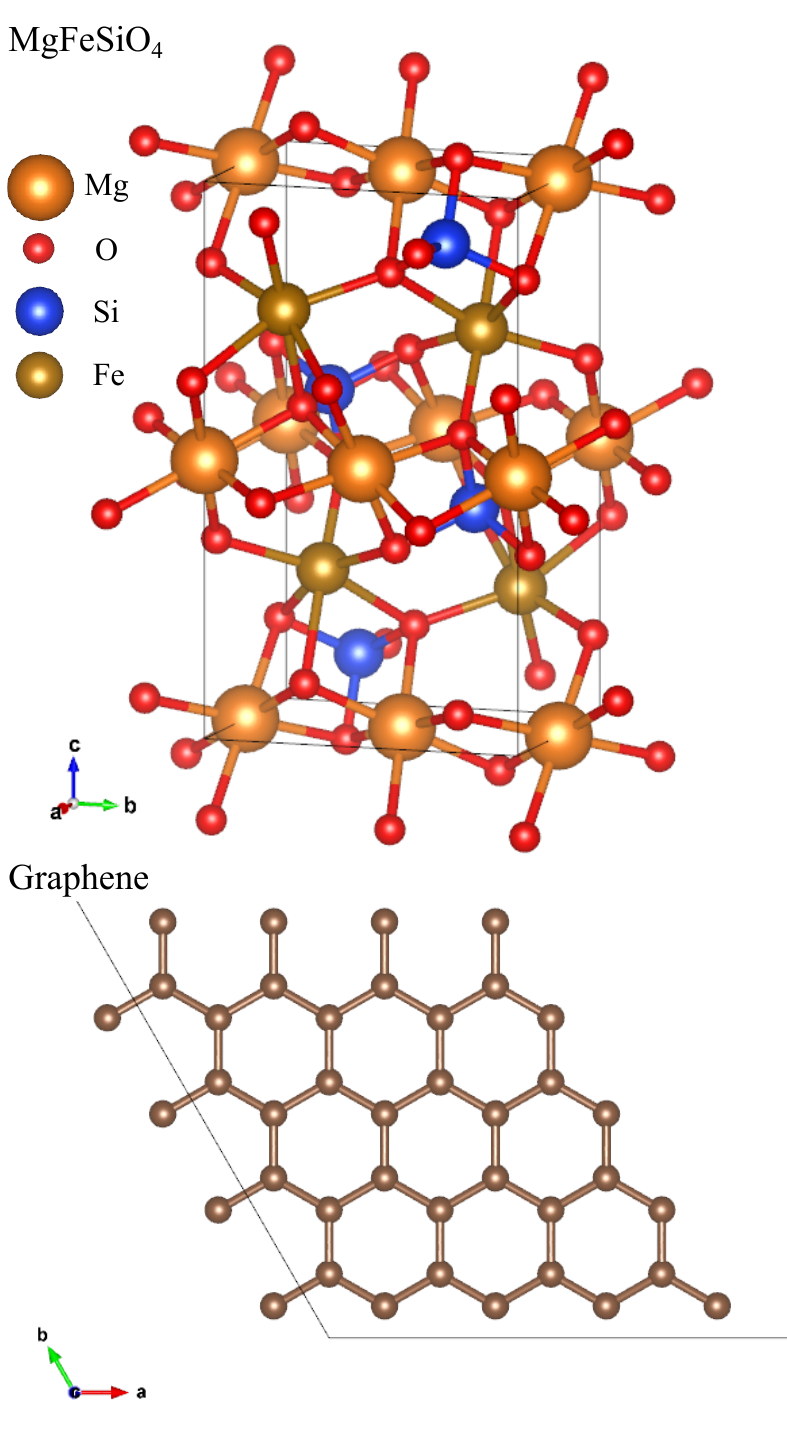}
  \caption{Structure of the model iron-bearing silicate in
    the \texttt{pnma} space group (\chem{MgFeSiO_4}, upper
    panel) and graphene as a proxy of graphites \citep[see
    also the Materials Project; ][]{MatProj}, both involved
    in binding energy calculations and KMC simulations.  }
  \label{fig:mgfesio4-graphene}
\end{figure}

\subsubsection{Iron-bearing Silicates: $\chem{MgFeSiO_4}$}

\begin{deluxetable}{ccc}
  \tablecolumns{3} 
  \tabletypesize{\scriptsize}
  \tablecaption{Binding energy of atoms in iron-bearing
    model silicates (as \chem{MgFeSiO_4}).
    \label{table:ene-mgfesio4} }
  \tablehead{
    \colhead{Concerned atom} &
    \colhead{Nearest neighbors} &
    \colhead{$\Delta E/\eV$}
  }
  \startdata
 Mg & 6O & 5.38 \\
    & 5O & 5.00 \\
    & 1O & 2.73 \\
 \hline
 Fe & 6O & 5.51 \\
    & 5O & 5.29 \\
    & 1O & 4.53 \\
 \hline
 Si & 4O & 10.22 \\
    & 3O & 8.11 \\
    & 1O & 6.30 \\
 \hline
 O & 2Mg + 1Fe + 1Si & 8.92 \\
   & 1Mg + 2Fe + 1Si & 8.61 \\
  & 2Mg + 1Si & 4.17 \\
  & 2Mg + 1Fe  & 6.91 \\
  & 2Fe + 1Si & 8.33 \\
  & 1Mg + 2Fe & 6.52 \\
  & 1Mg + 1Fe + 1Si & 8.11 \\ 
  & 1Mg & 1.47 \\
  & 1Fe & 3.11 \\
  & 1Si & 6.10 \\
  \enddata
\end{deluxetable}

\begin{figure}
  \centering
  \includegraphics[width=1.0\linewidth]
  {\figdir/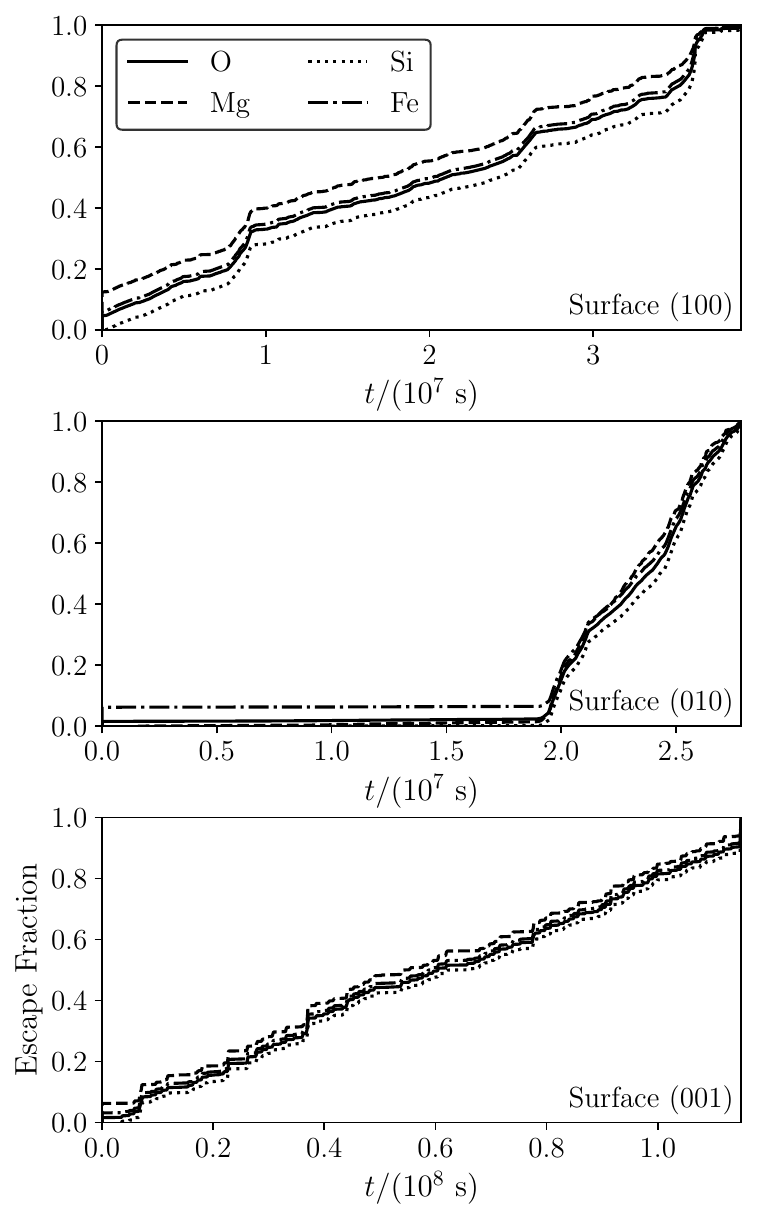}
  \caption{Similar to Figure~\ref{fig:kmc-mgsio3} but for
    model \chem{MgFeSiO_4} at $T = 1700~\K$. Note that the
    middle panel exhibits a stochastically delayed
    sublimaiton to $t\simeq 2\times 10^7~\s$ with the
    temporarily static structure after the first few atoms
    escape, whie the delay is not accounted for the
    calculation of sublimation rate.  }
  \label{fig:kmc-mgfesio4}
\end{figure}

The incorporation of iron into silicate lattices is
astronomically abundant and profoundly alters the material's
thermodynamic properties. Our simulations for
$\chem{MgFeSiO_4}$, which serves as a proxy of iron-bearing
silicates like olivine, reveal that it is significantly more
refractory than its magnesium-rich counterpart. The
calculated binding energies for Fe atoms in various
coordination environments are consistently higher than those
for Mg (especially at low coordination numbers), resulting
in substantially higher activation barriers for desorption
(see Figure~\ref{fig:kmc-mgfesio4}). Consequently, at any
given temperature, the sublimation rate for \chem{MgFeSiO_4}
is orders of magnitude lower than that of
\chem{MgSiO_3}. This implies that iron-rich silicate grains
can survive and be transported into hotter regions of the
disk much closer to the central star. Such finding suggests
a potential mechanism for the fractionation of dust
populations in the inner disk, where iron-poor silicates
sublimate earlier, potentially enriching the gas phase in
Mg, while iron-rich grains persist longer in solid form.

\subsubsection{Potassium-doped Silicates: The Fate of Alkali
  Metals}

The sequestration and release of alkali metals—potassium in
particular—are pivotal for the physical state of the
innermost protoplanetary disk.  Once liberated into the gas
phase, alkali atoms are expected to ionise quickly and
become the dominant charge carriers, tightening the coupling
between magnetic fields and gas and thereby fuelling the
MRI.  Current models diverge on how K is stored prior to
release: either as a surface layer adsorbed after formation
\citep{2006A&A...445..205I}, or as a species trapped inside
the silicate lattice itself \citep{2015ApJ...811..156D}.
Accurate MHD modelling of the innermost disk therefore
relies on quantifying the energetics of K–silicate binding.

We have performed DFT calculations of \chem{MgSiO_3}
lattices in which $\sim 1/40$ Mg sites are substituted by 
K atoms.  While fully coordinated K exhibits a modest (but
positive) binding energy, any under-coordinated K located at
or near the surface yields a negative binding energy
(Table~\ref{table:ene-mgsio3-k}).  The negative value
signifies a spontaneous, repulsive ejection of the K atom
into the gas phase—an outcome that is largely
temperature-independent.  Consequently, the disk is
continuously seeded with alkali ions, sustaining MRI-driven
accretion throughout the dust sublimation zone.  Because
this repulsive expulsion operates only when K atoms reside
within a few atomic layers of the surface, our results
favour the lattice-inclusion picture of
\citet{2015ApJ...811..156D}, while emphasising that surface
proximity is the decisive factor for release and must be
included in any consistent inner-disk chemical model.

\begin{deluxetable}{ccc}
  \tablecolumns{3} 
  \tabletypesize{\scriptsize}
  \tablecaption{Binding energy related to doped potassium
    atoms in model silicates \chem{MgSiO_3}.
    \label{table:ene-mgsio3-k} }
  \tablehead{
    \colhead{Concerned atom} &
    \colhead{Nearest neighbors} &
    \colhead{$\Delta E/\eV$}
  }
  \startdata
  K & 6O & 1.45 \\
    & 5O & $-2.35$ \\
    & 1O & $-1.18$ \\
  \hline
  O & 1K + 1Mg + 2Si & 6.71 \\
    & 1K + 2Si & 1.88 \\
    & 1K, 1Mg, 1Si & 1.98 \\
    & 1K & 2.15  
  \enddata
\end{deluxetable}

\begin{figure}
  \centering 
\includegraphics[width=1.0\linewidth]
{\figdir/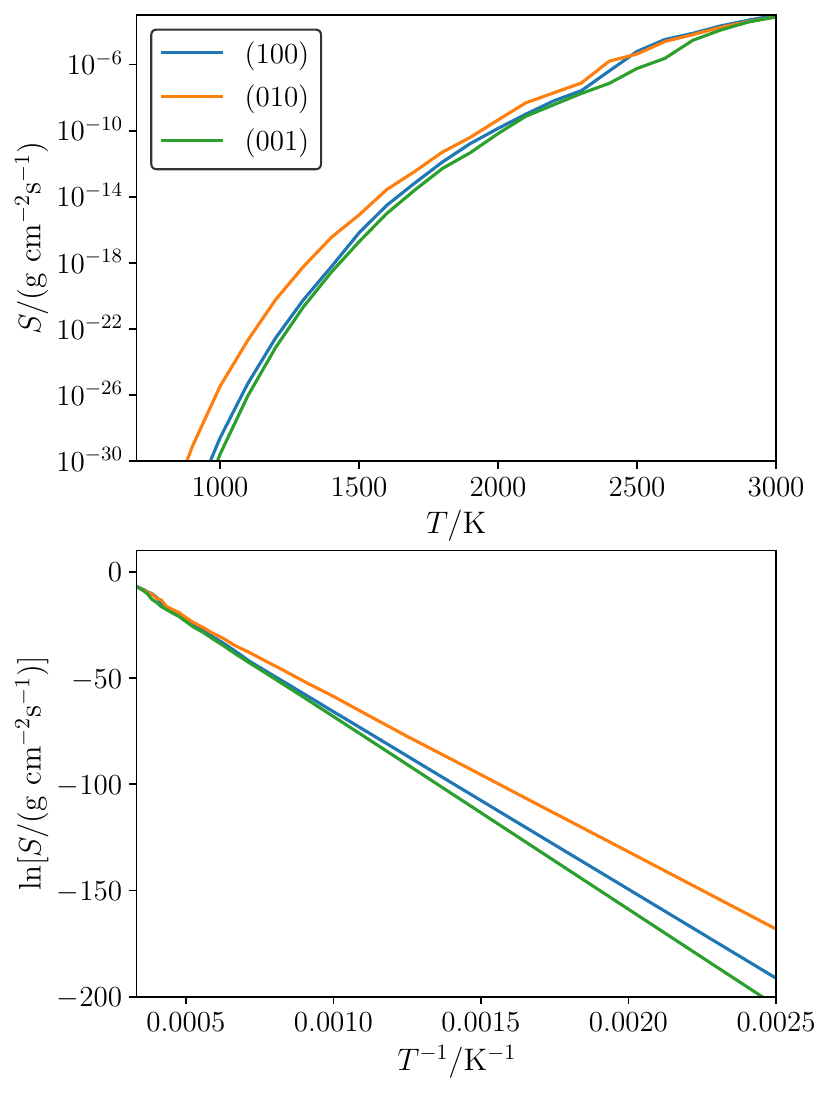}
\caption{Similar to Figure~\ref{fig:rate-mgsio3} but for
  iron-bearing silicate model \chem{MgFeSiO_4}. }
\label{fig:rate-mgfesio4}
\end{figure}

\subsubsection{Graphene for Carbonaceous Grains}

\begin{figure}
  \centering 
\includegraphics[width=1.05\linewidth]
{\figdir/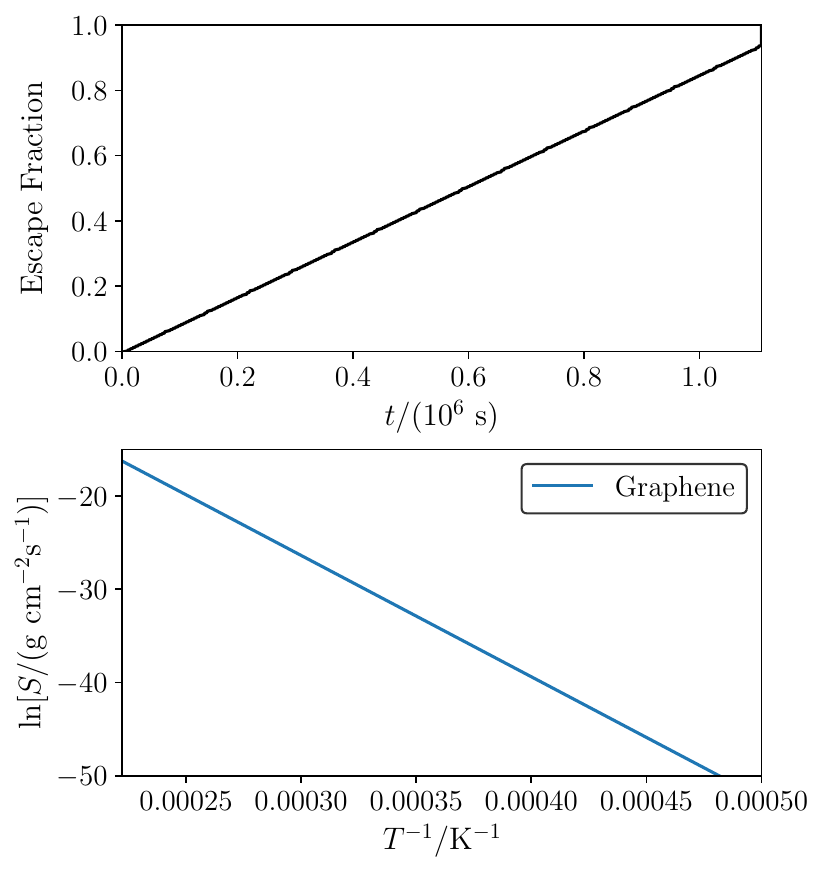}
\caption{Similar to Figures~\ref{fig:kmc-mgsio3} and
  \ref{fig:rate-mgsio3} but for graphene, as a proxy of
  graphites. Note that only one crystal surface is relevant
  in this case.}
\label{fig:rate-graphite}
\end{figure}

Carbonaceous dust represents a major population of solids
alongside silicates. We modeled the sublimation of graphitic
grains using a graphene-like structure, which could serve as
a reasonable proxy for both graphites and large polycyclic
aromatic hydrocarbons (PAHs). Our results, presented in
Figure~\ref{fig:rate-graphite}, demonstrate that pristine
graphite is exceptionally refractory, requiring temperatures
several hundred Kelvin higher than even \chem{MgFeSiO_4}
to achieve comparable sublimation rates. The strong covalent
bonding within the graphene layers creates immense
activation energies for carbon atom removal, making graphite
grains highly resilient in the inner disk.

However, this result requires important
context. Astronomical carbonaceous grains are not solely
composed of pure, crystalline graphite. They are often
amorphous and incorporate a significant fraction of volatile
components. These volatile organics will sublimate at much
lower temperatures, well before the graphitic backbone
begins to decompose. Therefore, while our simulation
accurately describes the sublimation of the most refractory
carbon component, the effective "sublimation front" for a
complex carbonaceous grain will be a broad zone dictated by
the loss of these weaker volatile coatings long before the
graphitic core sublimates. The detailed modeling of this
multi-stage process, involving the sequential sublimation of
different carbonaceous phases, is a complex but necessary
endeavor reserved for future work.

\section{Dust Survival in the Protoplanetary Disk Inner
  Regions} 
\label{sec:apps-dust-survival}

A direct and critical implication of our kinetic sublimation
calculations, detailed in \S\ref{sec:micro}, provides an
updated quantification of dust grain survival within the hot
inner regions of protoplanetary disks. Using simplified
models, semi-quantitative analyses could be helpful in
understanding the astrophysical impacts of non-equilibrium
sublimation kinetics.

\subsection{Survival Timescales of Stationary Grains}

\begin{figure} 
  \centering
  \hspace{-0.5cm} \includegraphics[width=1.05\linewidth]
  {\figdir/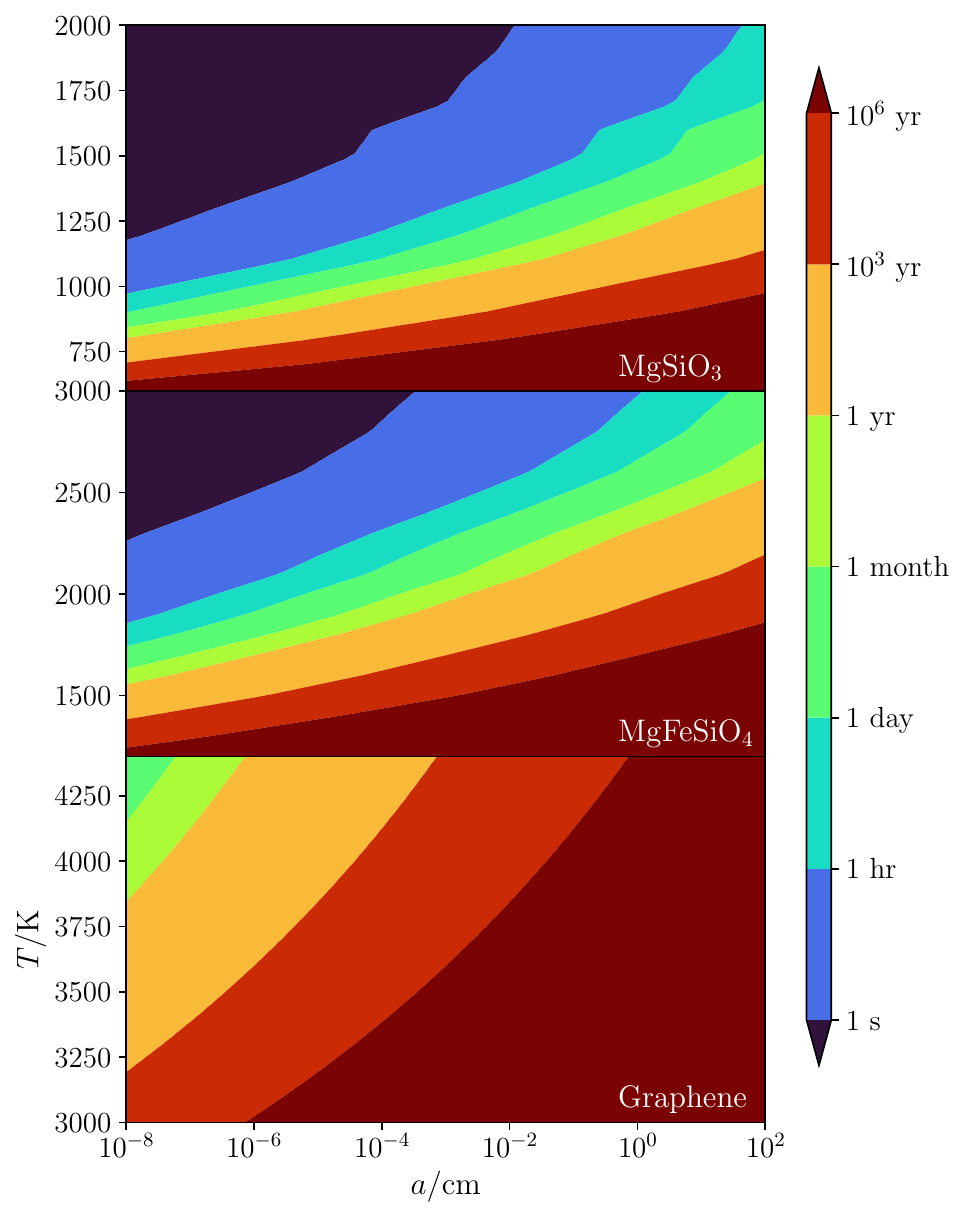}
  \caption{Survival timescales of dust grains with different
    components (indicated for each panel) at various
    temperatures ($T$, vertical axes) and grain sizes ($a$,
    horizontal axes), assuming spherical grains. }
    \label{fig:tau-survival}
\end{figure}

For a static disk model where dust grains are stationary and
experience no advective transport, our results indicate that
the survival timescale of a grain against sublimation is
inversely proportional to its radius, assuming grains of
different sizes are self-similar in composition and
structure. This size dependence arises because the
sublimation process is surface-area-controlled, while the
grain's total mass is volume-dependent. Assuming spherical
grains for this fiducial calculation, we present the derived
sublimation timescales as a function of temperature and
grain size in Figure~\ref{fig:tau-survival}. One may notice
from the figure that the contour corresponding to a survival
time of $10^6~\yr$, the typical lifetime of a protoplanetary
disk, lies at a remarkably low temperature of only
$T\sim 600-700~\K$, which is significantly lower than the
canonical $\sim 1200-1500~\K$ sublimation temperatures
derived from equilibrium thermodynamics. We nevertheless
emphasize that such discrepancy does {\it not} actually move
the sublimation front to $\sim 600~\K$. Instead, it must be
noted that the complete sublimation at low temperatures
could take excessively long period of time
($\gtrsim 10^6~\yr$), which underscores non-equilibrium
calculations (rather than instantaneous equilibrium phase
change), including inward advection replenishment to
maintain the dust grain persistance in the disk
interiors. In the meantime, the inclusion of Fe move the
survival time countours upwards by $\sim 600~\K$, preserving
solid materials at distances significantly closer to the
central star. Such differentiation caused by ingredients is
semi-quantitatively consistent with experimental results
\citep[e.g.][] {1994ApJ...421..615P, 2016JGRB..121.6384B},
which imposes a possible constraint on dust grain components:
existance of silicate dust grains at temperatures higher
than the \chem{MgSiO_3} survival conditions should be
iron-rich. Nevertheless, quantitative results should only be
obtained through comprehensive modeling of temperature
distributions in concerned astrophysical systems.

The location of the actual sublimation front in a realistic
disk is likely more complex. The static model provides a
foundational timescale, but it may be significantly altered
by the continual replenishment of dust grains via radial
accretion and mixing from cooler outer regions of the
disk. This dynamic replenishment could sustain a population
of dust grains inside the nominal kinetic sublimation zone,
implying that the observable ``sublimation front'' is not a
static sharp boundary, but a dynamic, smeared interface set
by the competition between inward transport and thermal
destruction.

\subsection{One-dimensional Advection Model}
\label{sec:apps-sub-1d}

To semi-quantitatively assess the impact of non-equilibrium
dust sublimation on the chemical and dynamical structure of
the inner protoplanetary disk, a simplified physical model
is developed to couple important channels of dust evolution
(including size distribution, radial transport, and thermal
destruction). A fundamental assumption of this simplified
model is that dust grains achieve a local equilibrium
between coagulation and fragmentation on timescales that are
short compared to those of radial drift and
sublimation. This justifies the use of a steady-state size
distribution, for which we adopt the canonical
Mathis-Rumpl-Nordsieck (MRN) form assuming spherical grains,
\begin{equation}
  \label{eq:mrn}
  \dfrac{\d n}{\d a} = \dfrac{n_0}{a_\max} \left(
  \dfrac{a}{a_\max}\right)^\xi\ ,\quad a\in (a_{\min}, a_\max)\ ,
\end{equation}
where $\xi$ is the power-law exponent (typically
$\xi = -3.5$), $a_\min$ and $a_\max$ are the minimum and
maximum grain radii, and $n_0$ is the normalization
parameter related to the grain number density $n_{\rm
  d}$. Such simplification permits semi-analytical
computations of the total dust mass density $\rho_{\rm d}$
(not to be confused with the the intrinsic material density
of solid materials of grains,
$\rho_{\rm s} \simeq 3~\g~\cm^{-3}$), defined as the dust
mass in unit overall space,
\begin{equation}
  \label{eq:def-rho_d}
  \begin{split}
    \rho_{\rm d} & = \rho_s \int_{a_{\min}}^{a_\max}\d a\ 
    \left(\dfrac{\d n}{\d a} \right)
    \left( \dfrac{4\pi a^3\rho_{\rm s}}{3} \right)
    \\
    & = \dfrac{m_\max n_0}{\xi + 4}
    \left[ 1 - \left( \dfrac{a_\min}{a_\max}\right)^{\xi+4}
      \right] \simeq \dfrac{m_\max n_0}{\xi + 4}\ ,
  \end{split}
\end{equation}
where $m_\max \equiv 4\pi \rho_sa_\max^3/3 $ is the mass of
a grain of radius $a_\max$, and the approximation arises
from $(a_\min/a_\max) \ll 1$. The thermal destruction of
dust due to sublimation is described by a phase transition
rate per unit volume, derived from the kinetic Monte Carlo
simulations presented in \S\ref{sec:micro-kmc},
\begin{equation}
  \label{eq:sublimation-rate}
  \begin{split}
  & \left( \dfrac{\partial \rho_{\rm d}}{\partial t}
    \right)_{\rm sub} = \int_{a_\min}^{a_\max} \d a\
    \left( \dfrac{\d n}{\d a} \right) 4\pi a^2 S(T)
  \\
  & \quad = \frac{3(\xi + 4)\rho_{\rm d}}{(\xi + 3)a_\min
    \rho_{\rm s}}
    \left[ 1 - \left( \dfrac{a_\min}{a_\max} \right)^{\xi + 3}
    \right] S(T)\ ,
  \end{split}
\end{equation}
where $S(T)$ is the temperature-dependent
sublimation rate, measured by the mass loss per unit surface
area taken from the KMC results (e.g.,
Figure~\ref{fig:rate-mgsio3}). The overall evolution of the
dust density is governed by the continuity equation that
incorporates both radial transport and sublimative mass
loss,
\begin{equation}
  \label{eq:continuity}
  \frac{\partial \rho_{\rm d}}{\partial t} +
  \nabla \cdot (\vec{v}
  \rho_{\rm d}) = - \left( \frac{\partial \rho_{\rm d}}
    {\partial t} \right)_{\rm sub}.
\end{equation}
Assuming an axisymmetric disk and purely radial transport
($\vec{v} = v_r \hat{r}$ for both dust and gas), the Minimum
Mass Solar Nebula (MMSN) model, where the gas surface
density follows
$\Sigma_{\rm g} =\Sigma_{\rm g,1}(r/\au)^{-1}$, relates the
radial velocity to the gas accretion rate by
$\dot{M}_{\rm acc} = 2\pi r \Sigma_{\rm g} v_r$. We seek a
steady-state solution
$\partial \rho_{\rm d} /\partial t = 0$, so that the
continuity equation of dusts then reduces to the logarithmic
form,
\begin{equation}
  \label{eq:1d-steady-log}
  \dfrac{\d \ln \rho_{\rm d}}{\d \ln r} = -2 - \dfrac{3(\xi +
    4)}{(\xi + 3) a_\max \rho_{\rm s} v_r}
  \left[ 1 - \left( \dfrac{a_\min}{a_\max} \right)^{\xi + 3}
  \right] S(T)\ .
\end{equation}
This form reveals that the dust density profile steepens
significantly with an increasing sublimation rate $S(T)$,
and, crucially, is inversely related to the radial accretion
velocity $v_r$.

For a high gas accretion rate $\dot{M}_{\rm acc}$, the
replenishment timescale is short. Inward-flowing dust from
cooler outer regions can move deeply into the hot inner disk
before being completely sublimated. This results in a broad,
spatially extended sublimation zone where dust and gas
coexist over a wide range of temperatures, significantly
inward of the nominal $\sim 1000~\K$ equilibrium
front. Conversely, in a disk with a low accretion rate,
radial replenishment is slow. The dust population is rapidly
destroyed upon reaching the hot inner region, leading to a
steeper density gradient and a sublimation front that more
closely resembles a sharp boundary, albeit still at a lower
temperature than the equilibrium prediction due to kinetic
inhibition. A broad sublimation zone, modulated by the
accretion rate, implies a more gradual release of refractory
elements into the gas phase. This directly affects the
spatial distribution of key gas-phase species and the
condensation sequence of materials during planet
formation. Furthermore, by setting the dust density profile,
the accretion rate indirectly governs the ionization degree
in the inner disk. Since dust grains are efficient charge
absorbers, a shallower dust gradient sustained by high
$\dot{M}$ could lead to a radially more extended laminar
accretion zone of low turbulence
\citep{2016ApJ...819...68X}, thereby influencing planet
migration and the overall disk evolution.

\begin{figure}
  \centering
  \includegraphics[width=1.02\linewidth]
  {\figdir/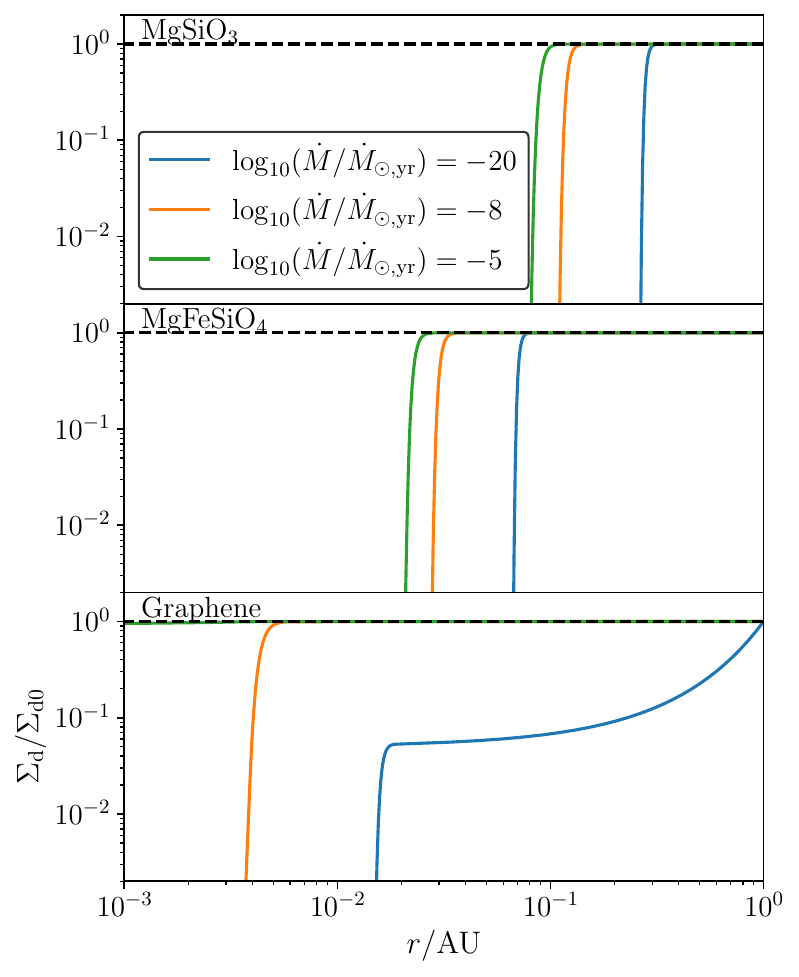}
  \caption{Radial distribution of dust surface density
    $\Sigma_{\rm d}$, normalized to the presumed value
    without sublimation $\Sigma_{\rm d0}$ , using simplified
    vertically unstratified accreting disk model assuming
    the MRN distribution is instantly reached
    (\S\ref{sec:apps-sub-1d}). Different accretion rates are
    distinguished by line colors indicated in the top panel,
    where $\dot{M}_{\odot,\yr}$ indicates a
    $1~M_\odot~\yr^{-1}$ accretion rate. The horizontal
    dashed lines in all panels indicate the $\Sigma_{\rm d}/
    \Sigma_{\rm d0} = 1$ condition for reference. }
    \label{fig:adv-simple}
\end{figure}

\subsection{Two-dimensional Model with Co-evolved Dust
  Evolution and Advection} 
\label{sec:apps-sub-2d}

In order to embed the sublimation rate calculation results
into protoplanetary disks more consistently, one must also
include advection profiles that is vertically stratified,
plus the coagulation and fragmentation processes alongside
with the sublimation. We hence composed the code
\jade{}\footnote{\url{https://github.com/wll745881210/JADE}}
in the Julia programming language to co-evolve these
relevant processes in 2.5D axisymmetric spherical polar
grids. The major physical mechanisms invloved are described in
what follows.

\begin{figure*}
  \centering
  \includegraphics[width=0.495\linewidth]
  {\figdir/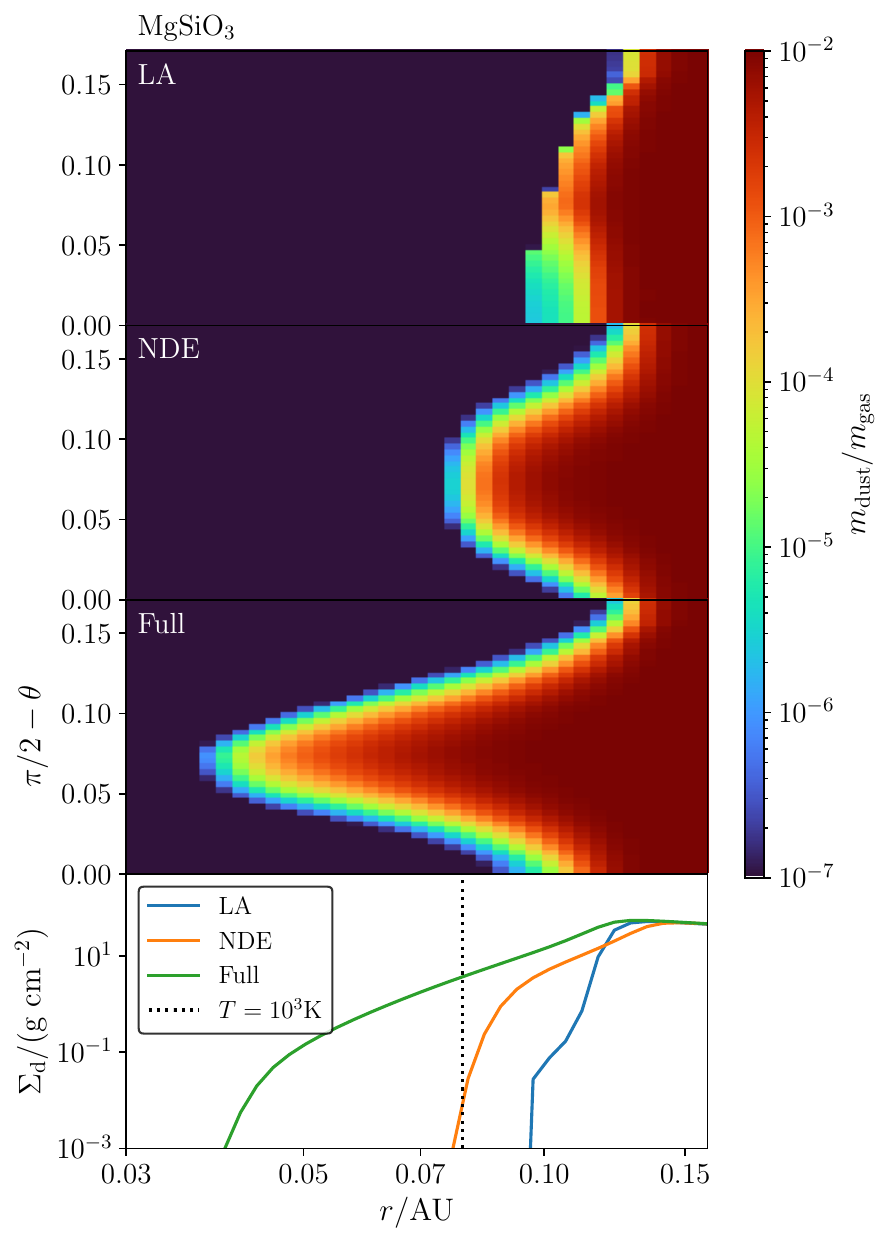}
  \includegraphics[width=0.495\linewidth]
  {\figdir/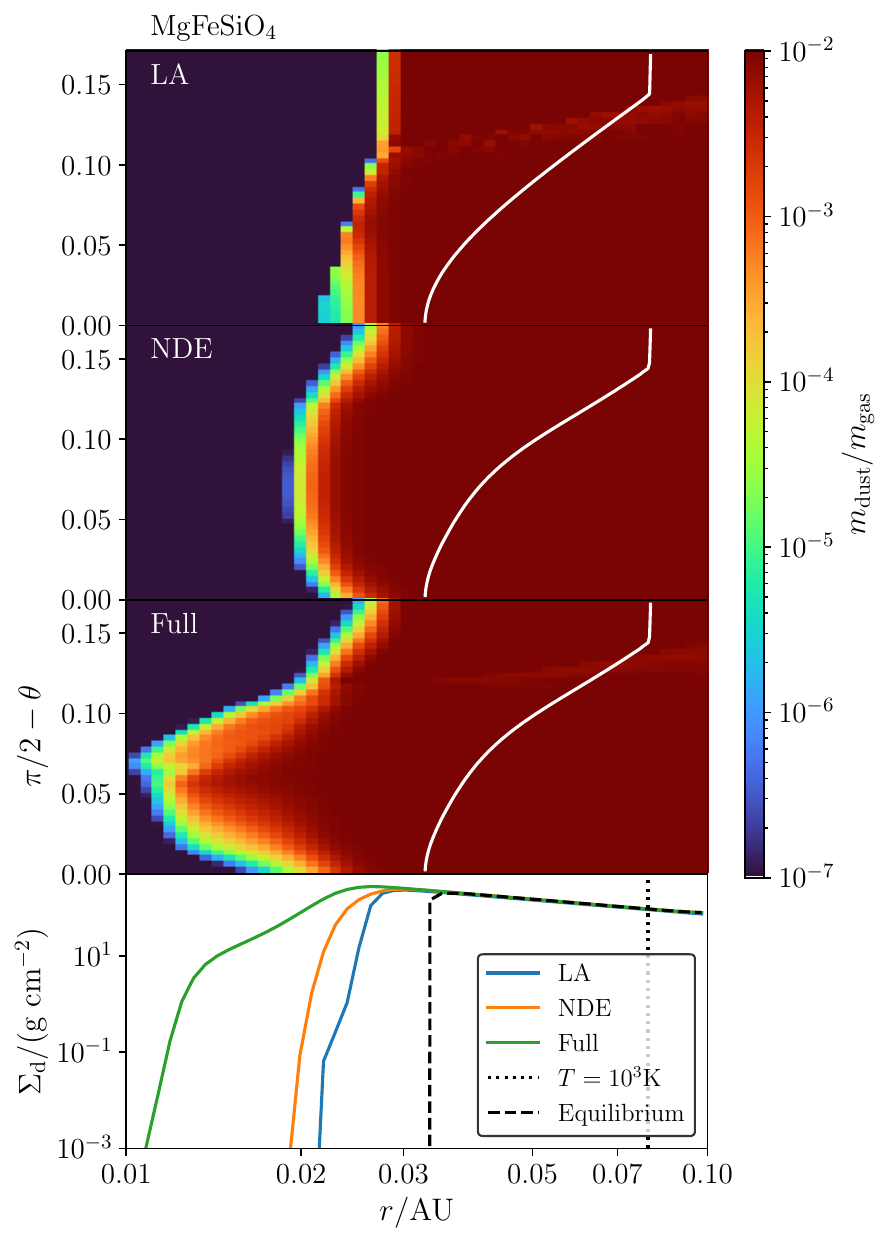}  
  \caption{Distributions of dust grain densities using
    \jade{} to calculate the advection and dust evolution
    simulationsly, for \chem{MgSiO_3} (left column) and
    \chem{MgFeSiO_4} (right column). In the top three panels
    in each column, Models LA (for ``low-accretion''), NDE
    (for ``no-dust-evolution''), and Full (model labels
    indicated on the upper-left corner in each panel), are
    presented in colormaps for dust-to-gas mass ratio
    ($m_{\rm dust}/m_{\rm gas}$) on the parameter planes
    subtended by the radius to the central protostar ($r$,
    in logarithmic scales) and the latitude
    ($\pi/2 -\theta$). The bottom panel in each column shows
    the vertically integrated dust surface density
    $\Sigma_{\rm d}$ through the concerned radii, in which
    the vertical dotted line indicates the radius at which
    $T=10^3~\K$ for reference. The equilibrium sublimation
    front of olivine (as iron-bearing silicates) are
    presented on the right column for comparisons using the
    criteria in \citet{1994ApJ...421..615P,
      2005A&A...438..899I}, as white contours in the
    colormaps and the black dashed line in the surface
    density plot. }
    \label{fig:prof-2d}
\end{figure*}

Figure~\ref{fig:prof-2d} presents the steady-state dust
surface-density profiles obtained with the \jade{} code, for
which the thermodynamic and kinematic fields are prescribed
as follows.  The gas density is given by
\begin{equation}
  \label{eq:2d-rho}
  \rho_{\rm g}(R,z)=\rho_{\rm
    g0}\left(\dfrac{R}{R_{0}}\right)^{\!\beta} 
  \exp\left(-\dfrac{z^{2}}{2h^{2}}\right),
\end{equation}
where the scale-height $h=c_{s}/\Omega$, and the
normalisation constants takes
$\rho_{\rm g0}=2.85\times 10^{14}~m_p~\cm^{-3}$ and
$R_{0}=1~\au$ ($m_{\rm p}$ is the proton mass).  We adopt
$\beta=-2.25$ so that the surface density scales as
$\Sigma\propto R^{-1}$ when the temperature follows
$T\propto R^{-1/2}$.  The sound speed is
$c_{s}=(\gamma\kb T/\mu)^{1/2}$ with adiabatic index
$\gamma=1.4$ and mean molecular weight $\mu=2.35$.  The
temperature profile is taken as $T(R)=T_0 (R/R_0)^{-1/2}$
with $T_{0}=280~\K$ at $R_{0}=1~\au$, appropriate for a
protostar of luminosity $L=3 L_{\odot}$ \citep[see
e.g.][]{1997ApJ...490..368C}.  Kinematically, the disk is
assumed to move only radially ($\vec{v} = v_r \hat{r}$),
with velocity
\begin{equation}
  \label{eq:2d-v}
  v_{r}(r,\theta)=v_{r0}\left[-\frac{(\theta-\theta_{0})^{2}}
    {2(\Delta\theta)_{v}^{2}}\right]^{-1/2},
\end{equation}
where we set $\theta_{0}=\pi/2-0.07$ and
$(\Delta\theta)_{v}=0.04$.  This Gaussian-like angular
dependence emulates the wind-driven accretion pattern
observed in the innermost regions of protoplanetary disks
\citep[see also, e.g.][]{2017ApJ...845...75B,
  2019ApJ...874...90W, 2024ApJ...972..142W}.  As long as
$\Sigma\propto R^{-1}$, the prescription above yields a
constant accretion rate whenever $v_{r0}$ is independent of
radius.

Adopting a fixed initial dust-to-gas mass ratio
$(m_{\mathrm{dust}}/m_{\mathrm{gas}})_{t=0}=10^{-2}$, we
evolve each model with \jade{} toward the steady states
illustrated in Figure~\ref{fig:prof-2d}. It must be noted
that these results should be considered as numerical
experiments with controlled parameters and profiles, which
should not be quantitatively compared to fully consistent
models (such as \citealt{2016ApJ...827..144F,
  2019A&A...630A.147F}, which coupled hydrodynamics, dust
evolution, and radiative transfer calculations with
equilibrium sublimation-condensation models). The ``Full''
models employ a surface radial velocity
$v_{r0}=10^{-1}~\km~\s^{-1}$, which is translated into a
typical $\dot{M}_{\rm acc}\simeq 10^{-8}~M_\odot~\yr^{-1}$
accretion rate, and include all dust evolution processes
described in Appendix~\ref{sec:jade-method}.  In contrast,
the LA (``low-accretion'') models reduce the velocity to
$v_{r0}=10^{-5}~\km~\s^{-1}$ (and $\dot{M}$ also reduces
proportionally), while the NDE (``no-dust-evolution'')
models suppress coagulation and fragmentation altogether.
In the Full runs, rapid surface accretion coupled with
efficient coagulation locks monomers into larger aggregates,
enabling a significant dust fraction to survive inside
$0.1~\au$ and producing a radially protruding bump near the
disk surface.  Conversely, the mid-plane accretes two–three
orders of magnitude more slowly; the sublimation front there
recedes outward, yielding a curved, bowl-shaped interface.
For both silicate compositions the transition is smooth: the
dust surface density $\Sigma_{\rm d}$ declines gradually by
two orders of magnitude over $\lesssim 0.1~\au$, roughly
three times the width of the steepest local gradient. This
is in contrast with the sharp sublimation fronts yielded
from equilibrium calculations, illustrated with the
comparisons for the \chem{MgFeSiO_4} model (note that the
equilibrium sublimation temperature profile generally
adopted by e.g. \citealt{1994ApJ...421..615P,
  2005A&A...438..899I} is evaluated for iron-bearing
olivine).  This broad, smeared edge naturally reproduces the
shallow $10~\micron$ silicate feature observed in T Tauri
disks \citep{2006ApJ...639..275K}.

The selection effect on grain sizes is clearly imprinted in
the steady-state size distributions.  Deep inside the disk
($r \gtrsim 0.3~\au$) the population largely obeys the MRN
law (see also Appendix~\ref{sec:jade-method} and
Figure~\ref{fig:coag-example}), $n(a)\propto a^{-3.5}$
between the simulated grain size range
$a_{\min}=0.1~\micron$ and $a_{\max}\sim 10^2~\cm$ using 20
grain size bins. In this region coagulation and
fragmentation are in local balance and temperatures are too
low for sublimation, so the full reservoir of small grains
is preserved.  Moving inward, temperatures exceed
$\sim 1200~\K$ and sublimation begins to preferentially
remove the smallest particles; because the sublimation rate
per unit mass scales as $\propto a^{-1}$ (surface-to-volume
ratio), sub-micron grains are rapidly depleted, leaving a
pronounced deficit at small sizes.  Consequently, comparison
with NDE models highlights the crucial role of coagulation:
in the absence of grain growth the sublimation front is
shifted outward by roughly a factor of two, as persistently
small grains remain far more vulnerable to thermal
destruction.

\section{Discussion and Summary}
\label{sec:discussion}

The detailed microscopic modeling of dust grains is not only
a refinement of equilibrium chemistry, but also an update in
the controlling process that sets the shape, location and
variability of the inner dust front in protoplanetary disks.
Non-equilibrium sublimation coupled with advection can smear
the dust front into a shallow slope, whose contribution in
shaping inner rims is comparable to (or even higher than)
the mechanisms elaborated in the full radiative transfer
calculations assuming equilibrium sublimation-condensation
of grains immersed in saturated vapors to model inner rims
\citep{2016ApJ...827..144F}. It is noted that dust grains
drift at speeds comparable to, but not identical to, the
gas. In addition, the disk ambients pressure is typically
considerably lower than the saturated vapor pressure at
temperatures that is relevant to sublimation processes.
Consequently, freshly sublimated vapour is continuously
advected away from the grain surface before saturation can
be established. This finding corroborates the
non-equilibrium framework advocated by
\citet{NAGAHARA19961445} and highlights the need for fully
kinetic chemical networks in inner-disk models.

\subsection{Dependence of the sublimation front on various
  factors}

When the radial drift timescales is comparable to
sublimation, the dust survial is kinetically
limited. Therefore, the radial survival length of a grain is
the product of its sublimation time-scale and the radial
drift speed. For \chem{MgSiO_3} at $T = 10^3~\K$, the
survival timescales for $\micron$-size dust grains is around
1 month, which is sufficient to travel $0.05~\au$ given
inward drift velocity $v_r = 10^{-1}~\km~\s^{-1}$.
Consequently, disks with higher accretion rates $\dot{M}$
(and hence larger $v_r$) admit dust far closer to the star,
producing a progressively shallower and more time-variable
sublimation front.  Conversely, in quiescent disks
($\dot{M}_{\rm acc} \ll 10^{-8}~M_\odot~\yr^{-1}$), the
front sharpens and recedes outward
(Figures~\ref{fig:adv-simple} and \ref{fig:prof-2d}).  The
dependence on $\dot{M}_{\rm acc}$ must be encoded in the
retrieval of inner-disk radii from infrared interferometry
\citep{2022A&A...658A..36K}. 

Revealed by the DFT binding energy calculations and
confirmed by the KMC simulations, iron-bearing silicates
(\chem{MgFeSiO_4}) are noticed to possess binding energies
$0.5-1~\eV$ higher than their Mg-rich counterparts
(Tables~\ref{table:ene-mgsio3} and
\ref{table:ene-mgfesio4}). Such difference translates into
sublimation rates suppressed by more than two orders of
magnitude at $T>1200~\K$ Iron-rich grains therefore survive
at temperatures $500-700$~K higher, and can be radially
transported to significantly smaller radii before
destruction (see also Figure~\ref{fig:prof-2d}). In case of
in-situ planet formation, inward-drifting solids are the
primary feed-stock for the formation of close-in rocky
planets, preferential survival of iron-rich grains naturally
biases the condensed material toward higher Fe/Mg ratios.
This offers a simple, kinetically driven explanation for the
elevated bulk iron fractions inferred for some rocky
exoplanets \citep{2015A&A...582A.112B}.

In real PPDs, the accretion flow is highly stratified:
surface layers accrete at $v_r\sim 0.1$~km~s$^{-1}$, while
the mid-plane moves an order of magnitude slower
(\S\ref{sec:apps-sub-2d}; see also
\citealt{2017ApJ...845...75B, 2019ApJ...874...90W,
  2024ApJ...972..142W}). Because sublimation is surface-area
limited, the front becomes bowl-shaped, sublimating first
high in the atmosphere and gradually ``peeling'' downward as
grains sediment.  The resulting iso-abundance lines are no
longer vertical walls, but curved surfaces whose
morphologies ratio controlled by the competition between
sedimentation speed to radial advection.  Capturing this
geometry requires vertically resolved simulations rather
than one-dimensional models. Interior to the dust
sublimation surfaces, small grains disappear first, followed
by larger ones.

After the release of alkali metals into the gas phase, the
resulting increase in ionized species triggers the
magneto-rotational instability (MRI) and switches the
accretion mode from laminar wind-driven to turbulent MRI
\citep{2009ApJ...701..737B, 2011ApJ...736..144B,
  2013ApJ...769...76B, 2013ApJ...767...30B}. Because MRI
torque scales with $B_z^2/\Sigma$, the accretion rate drops
precipitously when the surface density $\Sigma$ plummets
after dust sublimation.  Materials therefore pile up at the
transition radius, creating an inner rim whose location is
set by the competition between sublimation and MRI
activation rather than by a simple temperature criterion.
This rim is intrinsically time-variable on the local orbital
period, offering a natural explanation for the week-scale
NIR flickering observed by VLTI/GRAVITY
\citep{2021A&A...654A..97G, 2023A&A...669A..59G}.

\subsection{Future Works}

Building on the present {\it ab-initio} estimates of dust
survival, several avenues promise to extend the scope of
this work substantially.  An important caveat, admittedly,
is that the velocity field is prescribed rather than
self-consistently evolved.  After dust being removed in the
mid-plane due to slow radial drifts, the electron fraction
rises and ambipolar diffusion weakens, potentially
relocating the accretion layer downward, or even lead to
magnetic breaking and the subsequent growth of MRI. In the
meantime, the non-ideal MHD mechanisms could also cause
asymmetric accretion pattern, where the layer of accretion
may become one-sided \citep{2024ApJ...972..142W}. One must
also be aware of the heat production by MHD diffusivities
and heat transfer via radiation deep inside the disks that
could significantly reshape the disk thermal structures, as
part of the feedback processes regarding dust survival
\citep[see e.g.][] {2016ApJ...827..144F,
  2019A&A...630A.147F}.  Therefore, a consistent prediction
of the sublimation-front morphology requires embedding the
kinetic sublimation processes within radiative non-ideal MHD
calculations that include ambipolar diffusion, Ohmic
resistivity, and the Hall effect via consistent
non-equilibrium thermochemical calculations, including
simulations
\citep[e.g.][]{2017ApJ...845...75B,2019ApJ...874...90W} and
semi-analytic solutions \citep{2024ApJ...972..142W}.
Descriptions of the dispersal processes for the inner
protoplanetary disks are expected to be updated regarding
the overall structure, evolution, and planet-forming
potential.

Extending the framework to more complicated grains, such as
amorphous carbon and volatile organics, will allow us to
track the full refractory-to-volatile transition that shapes
infrared spectra across all evolutionary stages of disks and
envelopes. Our current graphite model neglects the volatile
organic mantle that sublimates first.  Future work will
employ a multi-component model (volatile organics,
refractory carbon, and amorphous silicate) to study the
sequential loss of coatings and the resulting spectral
evolution in the mid-IR
\citep[e.g.][]{2004A&A...413..571G}. More complete coverage
over the size and chemical conditions of {\it ab-initio}
modeling on PAHs could be helpful in constraining their
properties using the increasing number of mid-IR
observations on PAH features obtained via JWST
\citep[e.g.][] {2022ApJ...939...22Z, 2023ApJ...943....1Z,
  2023ApJ...943...60Z}.

Using the binding-energy catalogues (e.g.,
Tables~\ref{table:ene-mgsio3}, \ref{table:ene-mgfesio4} and
\ref{table:ene-mgsio3-k}), the same DFT--KMC--\jade{}
pipeline can be directly extended into a broader suite of
astrophysical applications.  One immediate extension is the
rapid sputtering and sublimation that occur when
interstellar grains encounter the shock waves of supernova
remnants.  The temperature-dependent sublimation rate $S(T)$
derived here provides an {\it ab-initio} estimate of grain
lifetimes against such shocks, thereby refining the
thermochemical evolution of the post-shock medium and the
subsequent star-formation channels, updating the classic
discussions of \citet{1978MNRAS.183..367B},
\citet{1998ApJ...503..247J}, \citet{2016A&A...589A.132B}.

By modifying the KMC algorithm to incorporate the stochastic
arrival, adsorption, and surface diffusion of atoms from the
gas phase, the same framework can be employed to simulate
condensation processes. This would allow us to assess the
fundamental efficiency of dust grain formation and,
crucially, to explore potential selection effects among
elemental components. For instance, could kinetic barriers
during condensation lead to the preferential incorporation
of certain atoms (e.g., Mg over Fe) into growing grains,
thereby fractionating the condensate from the gas phase?
This approach can be applied to model condensation into both
crystalline and amorphous solids, offering insights into the
initial conditions of dust populations that are later
delivered to PPDs.

Integrating such comprehensive picture of grain formation
and destruction into dynamic astrophysical simulations is
also a desired next step. A self-consistent model that
couples dust nucleation, growth, sublimation, and radial
transport would be transformative for studying dust-life
cycles across diverse environments. This includes the winds
of asymptotic giant branch (AGB) stars, where dust forms and
is ejected into the interstellar medium (ISM); molecular
clouds and star-forming cores, where dust grains act as
sites for chemical reactions and influence cloud collapse;
and the extreme environments around active galactic nuclei
(AGNs). In the vicinity of AGNs, intense X-ray irradiation
fields drive a complex and violent competition between dust
destruction and formation mechanisms—including sputtering,
sublimation, condensation, and radial drift—all occurring
on short dynamical timescales \citep{2007MNRAS.380.1172H,
  2020ApJ...892..149T, 2023A&A...676A..73G}. Since radiation
dominates the thermodynamics in these regions, the spatial
distribution of dust temperatures should be directly
imprinted onto the measurable radiative features of the
system. Therefore, we can expect that the balance between
solid silicate grain components and gas-phase species,
particularly for key elements like Fe, could be directly
tested. By comparing synthetic observables from our models
with actual near-infrared contiuum and mid-infrared features
(specifically the features of solid silicates such as the
$10~\micron$ and $18~\micron$ bands) and gas-phase emission
lines of Fe, one can perform a check on the predicted
non-equilibrium chemistry, and portray the variations of the
AGN appearance in the infrared consistently in terms of both
astrophysics and microphysics.

Finally, the highly metal-rich debris disks surrounding
polluted white dwarfs provide a unique and stringent testbed
for our kinetic data
\citep[e.g.,][]{2008AJ....135.1785J}. These systems are
believed to be the remnants of tidally disrupted
planetesimals, and their atmospheric compositions serve as a
direct probe of extrasolar planetesimal
chemistry. Reproducing the observed infrared and optical
features of these disks will require self-consistent
modeling that incorporates our precise binding energies and
sublimation rates. Successfully matching these observations
would provide strong validation of our approach and shed
light on the composition of rocky bodies in extrasolar
systems.

\begin{acknowledgments}
  This work is supported by the NSFC General Project
  12573067. The computational resources supporting this
  work are provided by the Kavli Institute for Astronomy and
  Astrophysics, Peking University.  We thank our colleagues:
  Xue-ning Bai, Jeremy Goodman, Xiao Hu, Di Li, Rixin Li,
  Ji-feng Liu, Feng Long, Satoshi Okuzumi, Kengo Tomida,
  Haifeng Yang, for helpful discussions and suggestions on
  the contents of the paper.
\end{acknowledgments}

\bibliography{dust_sub}
\bibliographystyle{aasjournal}

\appendix

\section{Dust Evolution Methods in JADE}
\label{sec:jade-method}

While the advection part in \jade{} is straightforward when
the drift velocity field is prescribed, its dust evolution
procedures need further discussions. The dust coagulation
and fragmentation processes are calculated by solving
integral-differential equations, similar to the methods
described in \citet{2024MNRAS.529..893L}, and elaborated in
what follows.

Collisions among single grains or aggregates can be
classified, by ascending kinetic energy, into four
qualitatively distinct regimes \citep{1997ApJ...480..647D}:
(1) sticking, (2) restructuring of existing aggregates to
form larger ones, (3) fragmentation of existing aggregates,
and (4) shattering accompanied by the rupture of chemical
bonds.  For practical purposes, regimes (1) and (2) are
merged into a single ``sticking'' category.  Regime (4)
requires extreme impact energies capable of driving
solid-phase shocks that break intramolecular bonds, a
condition rarely met in planet-forming environments; it is
therefore neglected.

\subsection{Coagulation and fragmentation}

Following the semi-analytic prescriptions laid out by
\citet{1997ApJ...480..647D} and \citet{2009A&A...502..845O},
which we adopt here for clarity and simplicity, we treat the
elementary building blocks of dust growth as identical,
spherical monomers of radius $a_{\m}$ and mass
$m_{\m}= 4\pi a_{\m}^{3}\rho_{\m}/3$, where $\rho_{\m}$
denotes the solid density of a monomer.  Aggregates are
regarded as agglomerations of these monomers.  Without loss
of generality we label the more massive aggregate by the
subscript ``$0$'' and the lighter one by ``$1$'';
equivalently, the label reflects the number of monomers
contained in each cluster.  Three-body collisions are
ignored.

\subsection{Breaking Energy and Elastic Parameters}

The pivotal physical quantity governing aggregation is
$E_{\br}$, the energy needed to break the contact between
two grains.  We adopt the expression proposed by
\citet{1997ApJ...480..647D}, refined with experimental data
from \citet{2000SSRv...92..265B}:
\begin{equation}
\label{eq:e_br}
E_{\br}\simeq 43\,
\gamma_{\eff}^{5/3}\,\mathcal{E}^{-2/3}\,\tilde{a}^{4/3}\ ,\
\mathcal{E}\equiv\left[
\frac{1-\nu_{1}^{2}}{\mathcal{E}_{1}}
+\frac{1-\nu_{2}^{2}}{\mathcal{E}_{2}}
\right]^{-1}\ ,
\end{equation}
where $\tilde{a}=a_{\m}/2$ is the reduced monomer radius,
$\mathcal{E}$ the reduced elastic modulus, and
$\mathcal{E}_{i}$ and $\nu_{i}$ denoting the Young's moduli
and Poisson ratios of the contacting materials. While
\jade{} allows users to adopt their own parameters, we adopt
the representative values
$\mathcal{E}\simeq 10^{11}{\rm\ dyn~cm}^{-2}$ and $\nu=0.32$
for carbonaceous grains, while
$\mathcal{E}\simeq 5.4 \times10^{11}~{\rm dyn\ cm}^{-2}$,
$\nu=0.17$ could be used for silicates
\citep{1997ApJ...480..647D}.  These elastic prescriptions
are approximate and cannot self-consistently describe grains
bearing only a few adsorbate layers; a more rigorous
treatment of grain elasticity is deferred to future work.

\subsection{Critical Energy for Sticking vs. Fragmentation}

When two aggregates collide, the threshold energy separating
sticking from fragmentation is estimated by
$E_{\crit}\simeq3\,\bigl[N_{c}(m_{0})+N_{c}(m_{1})\bigr]\,E_{\br}$
\citep[] [table 3]{1997ApJ...480..647D}, where $m_{0,1}$ are
the cluster masses.  The factor $3$ reflects the fact that
larger projectiles distribute impact energy over many
monomer contacts, whereas small ones cannot.  For a cluster
of mass $m$, the number of monomer contacts is approximately
$N_{c}(m)\simeq\bigl(m/m_{\rm m}\bigr)\,D_{\rm f}$, with
$D_{\rm f}$ the fractal dimension, taken as $D_{\rm f} = 1$
in this work. If the centre-of-mass kinetic energy $E_{k}$
is below $E_{\crit}$, the outcome is sticking, producing a
new cluster of mass $m=m_{0}+m_{1}$.  Recent studies
\citep{2023A&A...670L..21A} suggest that for large
aggregates $E_{\crit}$ may become independent of $E_{\br}$
because collective modes dominate over monomer bond rupture.
In the present work we retain eq.~\eqref{eq:e_br} for
clarity and postpone a detailed investigation of alternative
fragmentation modes.

\subsection{Fragmentation Distribution}

When $E_{k}\ge E_{\crit}$, we assume fragmentation of both
clusters.  Statistically, the fragment mass spectrum follows
a continuous distribution
\begin{equation}
\mathcal{F}(m;m_{0},m_{1})\propto m^{-q}\;,
\end{equation}
with $q=1.5$ \citep{2009A&A...502..845O}.  The upper mass
cut-off $m_{\max}=m_{0}$ (the larger colliding cluster) is
fixed by conserving mass and energy,

\begin{equation}
\label{eq:break-constrain}
\begin{split}
  \mathcal{N}_{m}\int_{m_{\m}}^{m_{\max}}\d m\;
  \Bigl(\frac{m}{m_{\m}}\Bigr)^{-q}m &= m_{0}+m_{1}\;,\\
  \mathcal{N}_{E}\int_{m_{\m}}^{m_{\max}}
  \Bigl(\frac{m}{m_{\m}}\Bigr)^{-q}\mathcal{S}(m)
  &=\bigl[\mathcal{S}(m_{0})+\mathcal{S}(m_{1})\bigr]+\frac{E_{k}}{\gamma_{\eff}}\;,
\end{split}
\end{equation}
where $\mathcal{S}(m)\approx4\pi a_{\m}^{2}(m/m_{\m})^{2/3}$
estimates the surface area.  The actual fragment
distribution is
\begin{equation}
\label{eq:break-pdf}
\mathcal{F}(m;m_{0},m_{1})=\min\!\bigl
\{\mathcal{N}_{m},\mathcal{N}_{E}\bigr\}\, 
\Bigl(\frac{m}{m_{\m}}\Bigr)^{-q}\;.
\end{equation}

\subsection{Differential-integral equation of local dust
  evolution}

With the sticking and fragmentation prescriptions above, the
temporal evolution of the dust mass distribution $\rho(m,t)$
is governed semi-analytically by
\begin{equation}
\label{eq:size-dist-evo}
\frac{\partial_{t}\rho(m)}{m}
=\iint\!\!\d m'\d m''\;
\frac{\rho(m')}{m'}\left[
\frac{\kappa(m',m'';m)}{2}\frac{\rho(m'')}{m''}
-\kappa(m,m';m'')\frac{\rho(m)}{m}
\right]\;,
\end{equation}
where $\rho(m)$ is the mass density per unit grain mass (not
to be confused with the solid density $\rho_{\m}$) and
$\kappa(m',m'';m)=\kappa_{\rm c}+\kappa_{\rm b}$ is the reaction
kernel. The coagulation and fragmentation kernels are
defined as, 
\begin{equation}
  \label{eq:growth-kernel}
  \begin{split}
    \kappa_{\rm c}(m',m'';m)
    & \equiv \sigma(m',m'')\,\delta(m'+m''-m)
      \int \d v\, v\,\mathcal{V}(v;m',m'')
      \Theta\!\bigl[v_{\crit}(m',m'')-v\bigr]\ ,\\
    \kappa_{\rm b}(m',m'';m)
    & \equiv \sigma(m',m'')\int\!\d
      v\,v\,\mathcal{V}(v;m',m'')\,
      \Theta\!\bigl[v-v_{\crit}(m',m'')\bigr]\,
      \mathcal{F}(m;m',m'')\ , 
  \end{split}
\end{equation}
with $\sigma=\pi(a_{0}^{2}+a_{1}^{2})$ the geometric
cross-section, $v_{\crit}\equiv\sqrt{2E_{\crit}/\mu}$ the
sticking threshold velocity ($\mu$ is the reduced mass), and
$\delta$, $\Theta$ the Dirac delta and Heaviside functions.
The relative velocity distribution $\mathcal{V}(v;m',m'')$
is commonly taken as a single-point distribution at the
collision speed $v_{\rm coll}$
\citep{2007A&A...466..413O,2012ApJ...752..106O}.  We relax
this simplification by adopting a Maxwell-Boltzmann
distribution whose mean is $v_{\rm coll}$, a choice
validated against semi-analytic studies
\citep{2014ApJ...792...69P}.  The collision speed itself
follows the prescriptions of \citet{2012ApJ...752..106O},
set by the local hydrodynamic conditions (gas density
$\rho_{\g}$, temperature $T$) and the turbulent viscosity
parameter $\alpha$.  Eq.~\eqref{eq:size-dist-evo} is
therefore applicable to dust aggregation in any
astrophysical environment.

\begin{figure}
  \centering
  \includegraphics[width=0.4\linewidth]
  {\figdir/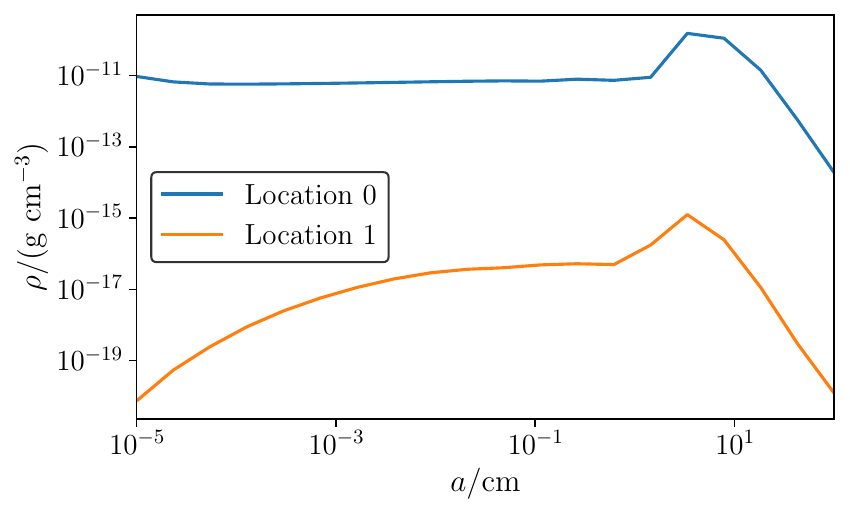}
  \caption{Example of dust evolution in the steady state of
    the Model Full for \chem{MgSiO_3}, presenting the dust size
    distributions (showing the total dust mass density $\rho$
    as a function of grain size $a$) for Locations 0  ($r =
    0.19~\au$ and $\pi/2 - \theta = 0$) and Location 1  ($r
    = 0.09~\au$ and $\pi/2 - \theta = 0$), respectively. }
    \label{fig:coag-example}
\end{figure}

An example of dust evolution calculation in the scenario of
an inner accreting disk is presented in
Figure~\ref{fig:coag-example}, comparing the dust mass
density distribution $\rho(m)$ at different locations in the
Model Full presented in the left column of
Figure~\ref{fig:prof-2d}. The curve showing the profile at
Location 0 ($r = 0.19~\au$ and $\pi/2 - \theta = 0$) agrees
with the MRN profile well at sizes smaller than the
coagulation bump (note that $\rho\propto a^{0.5}$ given
$\d n/\d a \propto a^{-3.5}$), while the curve for Location
1 ($r = 0.09~\au$ and $\pi/2 - \theta = 0$) exhibits a
significant reduction in small-size grains.

\end{document}